\documentclass[pra,twocolumn, showpacs,showkeys,secnumarabic,amssymb,amsmath,superscriptaddress,nofootinbib]{revtex4-1}
\usepackage{amssymb,amsmath,amsthm}
\usepackage{bbold}
\usepackage{color}
\usepackage{graphics}
\usepackage{graphicx}
\usepackage{enumitem}
\usepackage{bm}

\begin{document}

\title{A multipartite generalization of quantum discord}

\author{Chandrashekar Radhakrishnan}
\thanks{These authors contributed equally}
\affiliation{New York University Shanghai, 1555 Century Ave, Pudong, Shanghai 200122, China.}
\affiliation{NYU-ECNU Institute of Physics at NYU Shanghai, 3663 Zhongshan Road North, Shanghai, 200062, China.}
\affiliation{Laboratoire ESIEA Numérique et Société, ESIEA, 9 Rue Vesale, Paris 75005, France}

\author{Mathieu Lauri{\`e}re}
\thanks{These authors contributed equally}
\affiliation{Princeton University, ORFE Department, 98 Charlton St, Princeton, NJ 08540, USA}
\affiliation{New York University Shanghai, 1555 Century Avenue, Pudong, Shanghai 200122, China}

\author{Tim Byrnes}
\email{tim.byrnes@nyu.edu}
\affiliation{State Key Laboratory of Precision Spectroscopy, School of Physical and Material Sciences, East China Normal University, Shanghai, 200062, China.}
\affiliation{New York University Shanghai, 1555 Century Ave, Pudong, Shanghai 200122, China.}
\affiliation{NYU-ECNU Institute of Physics at NYU Shanghai, 3663 Zhongshan Road North, Shanghai, 200062, China.}
\affiliation{National Institute of Informatics, 2-1-2 Hitotsubashi, Chiyoda-ku, Tokyo, 101-8430, Japan.}
\affiliation{Department of Physics, New York University, New York, NY, 10003, USA.}

\begin{abstract}
A generalization of quantum discord to multipartite systems is proposed.  A key feature of our formulation is its consistency with the conventional definition of discord in bipartite systems. It is by construction zero only for systems with classically correlated subsystems and is a non-negative quantity, giving a measure of 
the total  non-classical correlations in the multipartite system with respect to a fixed measurement ordering.
For the tripartite case, we show that the discord can be decomposed into contributions resulting from changes induced by  non-classical correlation breaking measurements in the conditional mutual information and tripartite mutual information.
The former gives a measure of the bipartite  non-classical correlations and is a non-negative quantity, while the latter is related to the monogamy of the  non-classical correlations. 
\end{abstract}

\maketitle

%
%
%
\section{Introduction} 
One of the foremost aims of quantum information theory is to understand and quantify the various forms of quantum correlations.  Quantum correlations are ubiquitous in many areas of modern physics, ranging from condensed matter physics, quantum optics, high-energy physics, to quantum chemistry. They can be regarded as the most fundamental type of non-classical correlation which includes entanglement, EPR-steerable states, and non-local correlations \cite{adesso2016measures,ma2019operational}.  Much work has been done towards constructing resource theories \cite{dakic2012quantum,chitambar2016critical,veitch2014resource,gour2008resource,winter2016operational,madhok2013quantum} as well as understanding the 
operational relevance of information theoretic quantities 
\cite{PhysRevA.51.2738,cavalcanti2011operational,konig2009operational,devetak2006operational,giorda2010gaussian}. 

For bipartite systems, the best-known measure of the non-classical correlations is quantum discord (or discord for short) 
\cite{ollivier2001quantum,henderson2001classical}.  This is defined as the minimized difference between the quantum mutual information with and without a von Neumann projective measurement applied on one of the subsystems.  The role of the the projective measurement is to break the quantum correlations  (for simplicity, we henceforth use this term interchangeably with ``non-classical correlations'') between the subsystems, which results in a classically correlated state \cite{PhysRevA.84.042109,PhysRevA.82.034302}.  The intuition is that by comparing the mutual information before and after the breaking of quantum correlations, one can quantify the amount of quantum correlations in the original state with respect to the measured subsystem.   Quantum generalization of such entropies have focused on applications in quantum state distribution \cite{devetak2008exact,brandao2015quantum,ye2008quantum},
 optimal source coding \cite{yard2009optimal}, quantum information processing \cite{datta2008quantum,dakic2012quantum}, and simulation of classical channels with quantum side information \cite{luo2009channel,bennett2014quantum}.  Quantum discord has shown to be a powerful characterization tool for complex quantum states, such in quantum many-body systems \cite{werlang2010quantum,dillenschneider2008quantum}. 

For tripartite and larger systems, several generalizations of discord have been proposed. 
In Ref.  \cite{PhysRevA.84.042109} a symmetric multipartite discord was defined based on relative entropy and local measurements. Another definition of multipartite discord was provided in Ref. \cite{okrasa2011quantum}, as the sum of bipartite discords after making successive measurements.  An approach using relative entropy was defined in Ref. \cite{giorgi2011genuine} to define genuine quantum and classical correlations in multipartite systems. Ref. \cite{chakrabarty2011quantum} introduced the notion of quantum dissension defined as the difference between tripartite mutual information after a single measurement.  A distance-based approach was formulated in Ref. \cite{modi2010unified}, including a multipartite measure of quantum correlations \cite{modi2012classical,hu2018quantum,bera2017quantum}.  

 Surprisingly, a definition of quantum discord to multipartite systems that is consistent with the original 
bipartite definition of Refs. \cite{ollivier2001quantum,henderson2001classical} does not seem to exist. A reasonable set of properties \cite{PhysRevA.83.012312} possessed by such a measure include: (i) zero iff the state is a classically correlated state; (ii) a non-negative quantity; (iii) reduction to the 
 standard definition of discord for bipartite-like correlated subsystems.  In the original definition of quantum discord \cite{ollivier2001quantum,modi2012classical,hu2018quantum}, a classically correlated state is one such that the ``classical information is locally accessible, and can be obtained without perturbing the state of the system''.  This means that given a classically correlated state, there exists a measurement that can be performed such that the classical correlations can be recovered, without altering the density matrix.  For example, consider the state $ ( | 0 0 \rangle \langle 0 0 | + | 1 + \rangle \langle 1 + |) \otimes | 0 \rangle \langle 0 | /2  $.   As a tensor product of a bipartite zero discord state with a single qubit state, one expects that such a state to have 
zero tripartite discord, taking the first qubit to be the measured qubit.  Past works based on multipartite mutual information  \cite{okrasa2011quantum,PhysRevA.84.042109,giorgi2011genuine} give non-zero values for this state, either because of the type of measurements performed, or a symmetric definition. Meanwhile, quantum dissension \cite{chakrabarty2011quantum}, allows negative values which are not present with discord. For distance-based definitions \cite{modi2010unified}, one would not expect to obtain completely equivalent results due to the different measure used.  However, we note that Ref. \cite{modi2010unified} also uses a different notion of a classically correlated state to that of Ref. \cite{ollivier2001quantum}, and takes a non-zero value even for the bipartite component of the above state.

In this paper, a natural generalization of discord --- as originally defined in \cite{ollivier2001quantum,henderson2001classical} --- is made for multipartite systems.  Our definition satisfies all of the postulates of a multipartite discord (i)-(iii), thanks to the concept of conditional measurements which we introduce here. 
We further examine the entropy change to various mutual information quantities as a result of projective measurements, which leads to a method of decomposing the multipartite discord into various contributions.  This leads us to propose two more quantities based on mutual information, which measure the purely bipartite quantum correlations, and satisfy the properties (i)-(iii) as well as the monogamy of the quantum correlations in the tripartite system.

\section{Multipartite measurements}

Let us first start by reviewing the original definition of discord, which is defined as  \cite{ollivier2001quantum,henderson2001classical}
\begin{align}
D_{A;B}(\rho) =  \min_{ \Pi^A } \left[S_{B|\Pi^A} (\rho) - S_{B|A} (\rho) \right]
\label{eq:def_discord}
\end{align}
where the conditional entropy without measurement is defined $ S_{B|A} (\rho) = S_{AB} (\rho)  - S_{A} (\rho) $ \cite{cerf1997negative,cerf1999quantum,ollivier2001quantum},  where $ S_n(\rho) = - \text{Tr} \rho_n \log \rho_n $ is the von Neumann entropy for the (reduced) density matrix on the system labeled by $ n $.  The subsystem labels on the discord follow the notation such that a measurement is made on the label preceding the semicolon.  The conditional entropy with measurement is defined \cite{ollivier2001quantum}
\begin{align}
S_{B|\Pi^A} (\rho) = \sum_{j} p_{j}^A S_{AB} (\Pi_{j}^{A} \rho \Pi_{j}^{A}/p_{j}^A ) , 
\label{zurekcondent}
\end{align}
where $ \Pi_{j}^{A} $ is a one-dimensional von Neumann projection operator on subsystem $ A $ and 
$ p_j^A = \text{Tr} (\Pi_{j}^A \rho \Pi_{j}^A )  $ is its probability. 
The discord is zero if and only if there is a measurement such that $ \rho = \sum_{j}  \Pi_{j}^{A} \rho  \Pi_{j}^{A} $. 
The fact that one can measure one system and yet leave the state unchanged is a signal that there are no quantum correlations taking $ A $ to be the measured subsystem. 

In the above formulation, only one of the subsystems is measured.  For bipartite systems, this is sufficient since the correlations are only between two subsystems. 
We first generalize the bipartite discord to the case where both subsystems are measured. Although redundant for the bipartite case, understanding this will prove useful when generalizing discord to multipartite systems. In order to keep a consistent definition of discord, we seek a measurement for zero discord states such that $ \rho = \sum_{jk}  \Pi_{jk}^{AB} \rho  \Pi_{jk}^{AB} $.  Such a measurement can always be constructed according to the form \cite{Note1}
%
\begin{align}
\Pi^{AB}_{jk} = \Pi^A_j \otimes \Pi^B_{k|j}
\label{twomeasurements}
\end{align}
where $ \Pi^B_{k|j} $ is a projector on subsystem $ B $ that is conditional on the measurement outcome of $ A $ \cite{ollivier2001quantum}. The projectors satisfy $ \sum_k \Pi^B_{k|j} =\mathbb{1}^B, \sum_j \Pi^A_{j} = 
\mathbb{1}^A $.  As mentioned in the original work of Ref. \cite{ollivier2001quantum}, this would physically corresponds some classical communication from $ A $ to $ B $ being exchanged to modify the measurement on $ B $.  

Using this form of a measurement, we can then write an equivalent expression for the discord (\ref{eq:def_discord}), 
where measurements are made on both systems \cite{Note1}
\begin{align}
D_{A;B}(\rho)  =  \min_{\Pi^{AB}}  \left[  S_{B|A} ( \rho_{\Pi^{AB}} ) -S_{B|A}(\rho) \right] ,
\label{twomeasurementscheme}
\end{align}
where $ \rho_{\Pi^{AB}} = \sum_{jk} \Pi_{jk}^{AB} \rho \Pi_{jk}^{AB}  $ is the state after measurement. Here the optimization is performed over projective measurements of the type given in (\ref{twomeasurements}).  For example, the zero discord state $ ( |00\rangle \langle 00 | + |1+ \rangle \langle 1+ | )/2 $ has an optimal basis $ \Pi^{AB} \in \{ |00 \rangle \langle 00 |, |01 \rangle \langle 01 |, |1+ \rangle \langle 1+ |, |1- \rangle \langle 1- | \}$,   where 
$ |\pm\rangle = (|0\rangle \pm |1\rangle)/\sqrt{2}$. 
Without conditional measurements, it would be impossible to obtain consistent results with the conventional definition of discord  because the states 
$ | 0\rangle, | + \rangle $ are not orthogonal by themselves.

For multipartite systems with $N $ subsystems, in general $ N -1 $ local measurements will be necessary in order to break all the quantum correlations \cite{PhysRevA.84.042109,okrasa2011quantum}. In an analogous way to (\ref{twomeasurementscheme}) it is possible to equally make $ N $ measurements, but this is unnecessary and adds an extra overhead to the optimization, hence we consider $ N -1 $ measurements henceforth.  
For multipartite systems, each successive measurement is conditionally related to the previous measurement. The $N-1$-partite measurement is written
\begin{align}
\Pi^{A_1\dots A_{N-1}}_{j_1 \dots j_{N-1}} = \Pi^{A_1}_{j_1} \otimes \Pi^{A_2}_{j_2|j_1}  \dots \otimes \Pi^{A_{N-1}}_{j_{N-1}|j_1\dots j_{N-2}} ,
\label{conditionalmeasurement}
\end{align}
where the $ N $ subsystems are labeled as $ A_i $.  Here the measurements take place in the order $ A_1 \rightarrow A_2 \rightarrow \dots A_{N-1} $.

\section{Multipartite quantum discord}

We now show that the relevant quantity to be minimized in (\ref{eq:def_discord}) can be deduced by a simple procedure, which always ensures that the discord takes a zero value for measured states.  Evaluating the entropy of the measured system $ S_{AB} ( \rho_{\Pi^A}) $, we observe that this can always be decomposed as
\begin{align}
S_{AB}(\rho_{\Pi^A})  - S_A(\rho_{\Pi^A}) =  S_{B|\Pi^A} (\rho) ,
\end{align}
where $ \rho_{\Pi^{A}} = \sum_{j} \Pi_{j}^{A} \rho \Pi_{j}^{A}  $. The left hand side takes the form of conditional entropy $ S_{B|A}(\rho_{\Pi^A}) $  and all terms involve the measured subsystem $ A $.  The right hand side takes the form of (\ref{zurekcondent}), and is the average entropy of the unmeasured system $ B $.  
	If $ \rho_{\Pi^A}$ is replaced by a more general state $\rho$,  the equality does not hold.  
The comparison of the left and right hand side for a general state is then related to the degree of quantum correlations for the measurement performed on $ A $.

We can follow the same strategy to obtain a multipartite generalization of discord. Examining tripartite systems first, the total entropy of  $ S_{ABC} (\rho_{\Pi^{AB}})  $ can be decomposed to give \cite{Note1}
\begin{align}
S_{ABC}(\rho_{\Pi^{AB}}) - S_A(\rho_{\Pi^{AB}}) - S_{B| \Pi^{A}} (\rho_{\Pi^{AB}})
= S_{C|  \Pi^{AB}} (\rho) 
\label{tripartitedecomp}
\end{align}
where we have defined $ S_{C|  \Pi^{AB}} (\rho)  = \sum_{jk} p_{jk}^{AB} S_{ABC} ( \Pi^{AB}_{jk} \rho  \Pi^{AB}_{jk} / p_{jk}^{AB}) $ and $ p_{jk}^{AB} = \text{Tr} (\Pi_{jk}^{AB} \rho \Pi_{jk}^{AB}) $. 
Here, the left hand side contains terms which involve the entropy of the subsystems  $ A B $ that are measured, and the right hand side is the average entropy of the unmeasured system $ C $.  We thus define
\begin{align}
D_{A;B;C} (\rho) = & \min_{\Pi^{AB}  } \Big[  - S_{BC|A} (\rho) 
 + S_{B| \Pi^{A}} (\rho) + S_{C|  \Pi^{AB}} (\rho) \Big] 
\label{tripartitediscord}
\end{align}
as a tripartite generalization of discord, for the measurement ordering $ A \rightarrow  B $. This is a non-negative quantity, and by construction is zero for any post-measured state  e.g. for the state $ ( |00\rangle \langle 00 | + |1+ \rangle \langle 1+ | \otimes |0 \rangle \langle 0 |)/2 $ we have $ D_{A;B;C} (\rho) = 0 $.  Importantly, it is also true in the reverse direction, that $D_{A;B;C} (\rho)  = 0 $ implies that the state is of the form $ \rho_{\Pi^{AB}} $ \cite{Note1}.  The tripartite discord has the attractive property that it reduces to the standard bipartite discord when only bipartite quantum correlations are present: 
$ D_{A;B;C} (\rho^{AB} \otimes \rho^C)  = D_{A;B} (\rho^{AB}) $, $ D_{A;B;C} (\rho^{BC} \otimes \rho^A)  = D_{B;C} (\rho^{BC})  $, $ D_{A;B;C} (\rho^{AC} \otimes \rho^B) = D_{A;C} (\rho^{AC}) $ \cite{Note1}.

\begin{figure}[t]
\includegraphics[width=\linewidth]{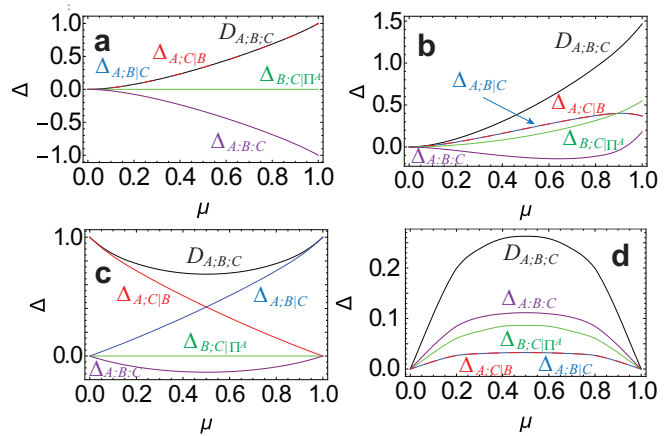}
   \caption{The tripartite quantum discord and its decompositions.   Definitions of quantities are given in (\ref{tripartitediscord}), (\ref{delta2}), (\ref{delta3}) and (\ref{delta2b}). 	The states are: (a) Werner-GHZ states $ \rho_{\text{W}} = \mu | \psi \rangle \langle \psi | + (1- \mu) \frac{\mathbb{1}}{8} $, where $ |\psi \rangle = (|000\rangle + |111\rangle)/\sqrt{2} $; (b) Werner-W states $ \rho_{\text{W}} $ defined the same as (a), but with $ |\psi \rangle = (|001\rangle + |010\rangle + |001\rangle)/\sqrt{3} $; (c) mixed Bell states $ \rho = \mu |\Phi^+_{AB} \rangle \langle \Phi^+_{AB} | + (1-\mu)|\Phi^+_{AC} \rangle \langle \Phi^+_{AC} |$, where $ |\Phi^+_{AB} \rangle = ( |000 \rangle + | 110 \rangle )/\sqrt{2} $, $ |\Phi^+_{AC} \rangle = ( |000 \rangle + | 101 \rangle )/\sqrt{2}$; (d) tripartite quantum correlated states $ \rho = \mu |000 \rangle \langle 000 | + (1-\mu)|+++ \rangle \langle +++ |$. 
The optimization is performed by minimizing the expression (\ref{tripartitediscord}) over all projection measurements $ \Pi_j \in \{ \cos \theta | 0 \rangle + e^{i\phi} \sin \theta | 1 \rangle , \sin \theta | 0 \rangle - e^{i\phi} \cos \theta | 1 \rangle  \} $  for the form (\ref{twomeasurements}), giving 6 parameters to optimize.  
}
\label{fig1}
\end{figure}

In Fig. \ref{fig1} we show several examples of the tripartite discord for various states. For the Werner states, we see that the tripartite discord generally follows a similar relation to bipartite discord, only diminishing to zero when $ \mu = 0 $, showing a similar behavior for the GHZ and W states. For a GHZ state, it is known that entanglement is present only for $ \mu > 1/5 $ \cite{pittenger2000note,dur2000classification,eltschka2012entanglement}, showing quantum correlations can be present even when entanglement is zero. The optimal
measurement (\ref{twomeasurements}) on the $ A $ subsystem is found to not necessarily coincide with the optimization for the bipartite discord between the $ A$ and $ BC $ subsystems.  This is because the expression (\ref{tripartitediscord}) contains contributions from other subdivisions. For Bell states the tripartite discord reduces to the bipartite values (Fig. \ref{fig1}(c)). The tripartite separable state shows quantum correlations as expected for any state that is not a product state  (Fig. \ref{fig1}(d)). This shows the non-convexity of the tripartite discord --- a property also shared by bipartite discord --- where a mixture of zero discord states can give a non-zero discord. 


The multipartite generalization can be performed by following the same logic. Evaluating the entropy of a $ N $-partite system measured using the conditional measurements (\ref{conditionalmeasurement}) we have the $ N $-partite discord
\begin{align}
& D_{A_1;A_2;  \dots ;A_N} (\rho)  =  \min_{ \Pi^{A_1 \dots A_{N-1} }} \Big[
- S_{A_2 \dots A_N|A_1 }(\rho) \nonumber \\
& +  S_{A_2| \Pi^{A_1} } (\rho)  \dots + S_{A_N|  \Pi^{A_1 \dots A_{N-1} } }  (\rho) \Big] 
\label{multipartitediscord}
\end{align}
for the measurement ordering $ A_1 \rightarrow A_2 \rightarrow \dots A_{N-1} $. Here 
 we have defined $ S_{A_k|\Pi^{A_1 \dots A_{k-1}}} (\rho) = 
\sum_{j_1 \dots j_{k-1}} p_{\bm{j}}^{(k-1)}
S_{A_1 \dots A_k}( \Pi^{(k-1)}_{\bm{j}}  \rho  \Pi^{(k-1)}_{\bm{j}}  / p_{\bm{j}}^{(k-1)} )$
with $ \Pi^{(k)}_{\bm{j}} \equiv \Pi^{A_1 \dots A_{k}}_{j_1 \dots j_{k}} $,  
$  p_{\bm{j}}^{(k)}  = \text{Tr} (  \Pi^{(k)}_{\bm{j}}  \rho  \Pi^{(k)}_{\bm{j}}  ) $. 
This is again a non-negative quantity, and reduces to lower order discords for states that have classically correlated subdivisions 
\cite{Note2} 
For a single $m$-dimensional system, the number of parameters to specify a projector is $ m(m-1) $ \cite{rau2018calculation}.  For $ N $ qubits, there are a total of $ \sum_{n=1}^{N-1} m^{n-1} $ local projectors in (\ref{conditionalmeasurement}), giving a total of $ m^N - m $ parameters to optimize in the discord (\ref{multipartitediscord}). 

\begin{figure}[t]
\includegraphics[width=\linewidth]{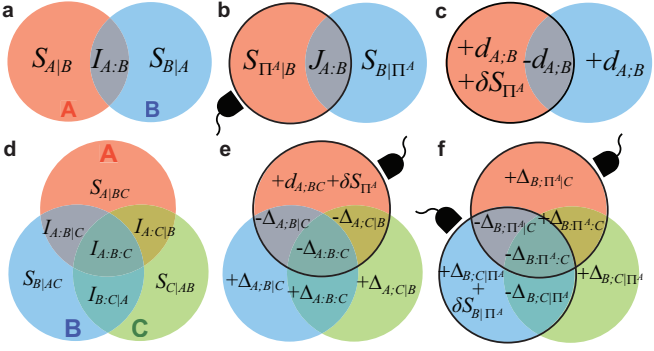}
\caption{Distribution of various entropies during a measurement in bipartite and tripartite systems. (a) and (d) The initial state before the measurement. (b) The final state after a measurement on subsystem $ A $.  The thick outline indicates the measured system.  (c) and (e) The change in entropy after a measurement on $ A $.  (f) The change in entropy after measuring both $ A $ and $ B $. Bipartite entropy contributions are defined according to $ I_{A:B}(\rho) = S_A(\rho) + S_B(\rho) - S_{AB}(\rho) $,  $ J_{A:B}(\rho) =  S_B(\rho) - S_{B| \Pi^{A}} (\rho)  =  I_{A:B} (\rho_{\Pi^{A}}) $, $  	S_{ \Pi^{A} |B}(\rho)  \equiv S_{AB} (\rho_{\Pi^A}) - S_B( \rho_{\Pi^A}) $, $ \delta S_{\Pi^A} (\rho) = S_A (\rho_{\Pi^A}) - S_A (\rho)  $,  $ d_{A;B}(\rho) =  S_{B|\Pi^A} (\rho) -S_{B|A} (\rho) $.  Tripartite entropy contributions are defined by $ I_{A:B|C} (\rho)  = S_{A|C} (\rho)  - S_{A|BC} (\rho) $, $ I_{A:B:C} (\rho) = I_{A:C}(\rho) - I_{A:C|B}(\rho) $, $ \delta S_{B|\Pi^A} (\rho) = S_{B|A} (\rho_{\Pi^{AB}}) - S_{B|A} (\rho_{\Pi^A}) $, $ \Delta_{B; \Pi^A |C} ( \rho) =  J_{A:B|C} (\rho) - K_{A:B|C} (\rho) $, $ \Delta_{B; \Pi^A ;C} ( \rho) =  J_{A:B:C} (\rho) - K_{A:B:C} (\rho) $. }
\label{fig2} 
\end{figure}

\section{Quantum discord as an entropy flux}

The multipartite generalization of discord gives a quantification of the total quantum correlations in the system with respect to a particular measurement ordering.  In a multipartite system, it is desirable to identify exactly where the quantum correlations exist in the system, to see the contributions between subsystems.  Before examining the multipartite case, it is interesting to revisit the bipartite case first.  The quantum correlation breaking measurement causes a pattern of entropy flux through the system. 
The entropy contributions before and after the measurement can be written as 
given in Fig. \ref{fig2}(a) and (b), where the same definitions of the entropies is used throughout except the state changes from $ \rho $ to $ \rho_{\Pi^A} $ \cite{Note1}. The entropy change for the three contributions are shown in Fig. \ref{fig2}(c).  We see that the measurement causes the mutual information to decrease by an amount equal to the discord, and the conditional entropies increase by the same amount.  The conditional entropy for the measured system $ A $ also increases by a local contribution $ \delta S_{\Pi^A} (\rho) $, since a measurement is applied on this subsystem.  This has the interpretation that the entropy corresponding to the quantum correlations are redistributed into subsystems $ A $ and $ B $ separately, since the measurement destroys this for the mutual information.

For tripartite systems, a similar redistribution of entropies occur.
The measurement (\ref{twomeasurements}) can be performed in two steps, first performing a measurement on $ A $, then conditionally performing another measurement on $ B $.  The initial distribution  is shown in Fig. \ref{fig2}(d), which changes to Fig. \ref{fig2}(e) after the first measurement. We define the measured version of the conditional mutual information $ I_{A:B|C} $ and tripartite mutual information $ I_{A:B:C} $ according to \cite{chakrabarty2011quantum,okrasa2011quantum,PhysRevA.84.042109}
\begin{align}
	J_{A:B|C} (\rho) & =  I_{A:B|C} (\rho_{\Pi^{A}} ) 	\label{eq:defJI} \\
	K_{A:B|C} (\rho) & =  I_{A:B|C} (\rho_{\Pi^{AB}} ) 	\label{eq:defKI}
\end{align}
and similarly for the remaining quantities  (mutual information is denoted with a colon). For a classically correlated state  the above definition ensures $I_{A:B|C}=J_{A:B|C}=K_{A:B|C}$, but more generally these quantities are not equal.  This naturally leads us to define various contributions to the entropy change as a result of the measurement.  After one measurement, the conditional mutual information changes by an amount
\begin{equation}
\label{delta2}
\begin{split}
\Delta_{A;B|C} (\rho)  & \equiv  I_{A:B|C} (\rho) - J_{A:B|C} (\rho) 
\\
&= d_{A;BC} (\rho) - d_{A;C} (\rho),
\end{split}
\end{equation}
which we call the {\it conditional tripartite discord}, and can be interpreted as the bipartite like quantum correlations in the system.  We may similarly define  $ \Delta_{A;C|B} \equiv  I_{A:C|B} (\rho) - J_{A:C|B} (\rho) $, where $ d_{A;C}(\rho) =  S_{C|\Pi^A} (\rho) -S_{C|A} (\rho) $ is the argument to be minimized for the bipartite discord.  This is a non-negative quantity $ \Delta_{A;B|C} (\rho),  \Delta_{A;C|B} (\rho) \ge 0 $, and reduces to the bipartite discord without the minimization: $ \Delta_{A;B|C} (\rho^{AB} \otimes \rho^C) = d_{A;B} (\rho^{AB}) $ \cite{Note1}.  

Similarly for the tripartite mutual information we define
\begin{align}
\Delta_{A:B:C}(\rho)  &\equiv  I_{A:B:C} (\rho) - J_{A:B:C} (\rho)  \nonumber \\
& = d_{A;B} (\rho) + d_{A;C} (\rho) - d_{A;BC} (\rho) .  
\label{delta3}
\end{align}
This can take positive or negative values \cite{chakrabarty2011quantum}.  The fact that this can be negative is not entirely surprising from the point of view that even classically, the tripartite mutual information can be negative. From the decomposition into discords, it is evident that this is a monogamy quantity, giving a  negative value for monogamous and  positive value for polygamous quantum correlations \cite{prabhu2012conditions}.   

Figure \ref{fig2}(e) shows the changes in the entropy after a measurement on $ A $ \cite{Note1}. We see that the entropy changes follow an analogous structure to the bipartite case (Fig. \ref{fig2}(c)).  The three contributions to the entropy $ \Delta_{A;B|C} , \Delta_{A;C|B}, \Delta_{A:B:C} $ are ``extruded'' to the unmeasured parts of the system.  The total of the three parts is equal to the conventional bipartite discord
\begin{align}
d_{A;BC} (\rho) = \Delta_{A;B|C} (\rho)  + \Delta_{A;C|B} (\rho)+ \Delta_{A:B:C}  (\rho)
\end{align}
which, combined with a local entropy increase $ \delta S_{\Pi^A} $, is also the increase in the conditional entropy of $ A $.  

After an additional measurement on $B$, a similar pattern emerges, except that the entropy shifts are in the direction of $ CA$ and $ CB $ instead of $ AB $ and $ AC $ as before.  Changes in the conditional mutual and tripartite mutual information are defined similarly to (\ref{delta2}) and (\ref{delta3}).  The most interesting of these terms is 
\begin{align}
\Delta_{B;C|\Pi^A} (\rho)  & \equiv  J_{B:C|A} (\rho) - K_{B:C|A} (\rho)  \nonumber \\
& = d_{B;\Pi^A C} (\rho) - d_{B;\Pi^A} (\rho) ,
\label{delta2b}
\end{align}
which is conditional discord after the measurement of $ A$, and is also non-negative:  $ \Delta_{B;C|\Pi^A} (\rho) \geq 0$  \cite{Note1}.  In addition to the similar pattern of entropy changes, there is again a local entropy contributions on subsystem $ B $.

The above definitions allow us to write the generalized discord (\ref{tripartitediscord}) in an equivalent form
\begin{align}
D_{A;B;C} (\rho) = & \min_{\Pi^{AB}} \Big[ 
\Delta_{A;B|C}  (\rho) + \Delta_{A;C|B} (\rho) + \Delta_{B;C|\Pi^A} (\rho)  \nonumber \\
& + \Delta_{A:B:C} (\rho)\Big] .
\label{tripartitediscroddecomp}
\end{align}
The tripartite discord can thus be equivalently viewed as the sum of all conditional discords and the change in the tripartite mutual information.  

This decomposition allows us to attribute various contributions of the total discord to various parts of the system. Figure \ref{fig1} shows the decompositions of the multipartite discord into various components.  For the Werner-GHZ state we see that the conditional discords between $ AB $ and $AC $ take the values $ \Delta_{A;B|C} = \Delta_{A;C|B} = D_{A;B;C} $, showing that bipartite quantum correlations exist between within the GHZ state, prior to a measurement on $ A $.  Meanwhile, the remaining conditional discord is $ \Delta_{B;C|\Pi^A} = 0 $ due to all quantum correlations (and hence entanglement) collapsing to zero after the measurement on $ A $ is made. The monogamous nature of the GHZ is verified with the change in the triparite mutual information, giving a negative value $ \Delta_{A:B:C} =  - D_{A;B;C} $.  For the Werner-W states, the conditional discord for all three pairings take non-zero values, since the measurement on $ A $ does not completely break the quantum correlations between $ BC $.  It is well known
that the tripartite quantum systems can be divided into these two classes, 
which are not related to each other via local operations and classical communication
\cite{dur2000three}.  Interestingly, the monogamy swaps sign from polygamous to monogamous behavior at lower purities. A similar effect was also found using a different measure in Ref. \cite{prabhu2012conditions}. For the bipartite states
in Fig. \ref{fig1}(c), the conditional discords reduce to the bipartite discords at $ \mu = 0,1 $.  Finally, for the tripartite correlated state all quantities are positive (Fig. \ref{fig1}(d)).

\section{Conclusions} 
We have introduced a generalization of discord for tripartite (\ref{tripartitediscord}) and multipartite (\ref{multipartitediscord}) states.  
One of the main features of our approach is the use of conditional measurements.  The conditioning is essential to take into account of all classical correlations 
that may exist between subsystems.  Viewing the measurements as an operation to break the quantum correlations, optimizing over all 
such measurements allows one to recover the purely quantum contribution.
 We note that there is an obvious asymmetry due the fixed ordering of the measurements, which is also present in the original definition of the bipartite discord.  While symmetric definitions of discord exist such that there is no dependence upon the choice (and order) of measured subsystems \cite{PhysRevA.84.042109,soares2010nonclassical}, here we take the point of view that we wish to have a definition consistent with  the most commonly used definition of bipartite discord as defined in Refs. \cite{ollivier2001quantum,henderson2001classical}.
This asymmetry has similarities with quantum steering which also considers a measurement on part of a system \cite{cavalcanti2016quantum}.  The aims are somewhat different in that for discord, it is to minimize the disturbance due to measurement rather than compare to a local hidden state theory \cite{he2015classifying}. Some applications, such as one-way quantum computing \cite{raussendorf2001one}, have a definite ordering of measurements which makes our multipartite discord naturally compatible.  By identifying the various contributions to the terms which makes up our definition of tripartite discord in terms of conditional entropies, we provide an exact decomposition (\ref{tripartitediscroddecomp}). The contributions give a definition of a conditional discord which characterizes the bipartite correlations $ \Delta_{A;B|C}, \Delta_{A;C|B}, \Delta_{B;C|\Pi^A} $  in a tripartite system, as well as a quantity related to the monogamy of quantum correlations $ \Delta_{A:B:C} $.   Similar decompositions can be made for the multipartite system, which we leave as future work.  

\section*{Acknowledgements}
We thank G. Adesso for illuminating discussions. This work is supported by the Shanghai Research Challenge Fund; New York University Global Seed Grants for Collaborative Research; National Natural Science Foundation of China (61571301,D1210036A); the NSFC Research Fund for International Young Scientists (11650110425,11850410426); NYU-ECNU Institute of Physics at NYU Shanghai; the Science and Technology Commission of Shanghai Municipality (17ZR1443600); the China Science and Technology Exchange Center (NGA-16-001); and the NSFC-RFBR Collaborative grant (81811530112).


\appendix

\section{Definitions and notations}
For convenience, we gather here some useful definitions and notations.

\noindent
\textbf{Projectors. }
 We denote by $ \Pi_{j}^{A} $ a one-dimensional von Neumann projection operator on subsystem $ A $  and by
$ p_j^A = \text{Tr} (\Pi_{j}^A \rho \Pi_{j}^A )  $ its probability. We use the notations
\begin{align}
	\rho_j &= \Pi_{j}^A \rho \Pi_{j}^A / p_j^A,
	\\
	\rho_{\Pi^A} &= \sum_j p_j \rho_j.
\end{align}
For measurements over $AB$, we use the notation 
\begin{align}
	\Pi^{AB}_{jk} = \Pi^A_j \otimes \Pi^B_{k|j}
\label{supp:twomeasurements}
\end{align}
where $ \Pi^B_{k|j} $ is a projector on subsystem $ B $ that is conditional on the measurement outcome of $ A $. For the associated probability and resulting state, we use the notations $p^B_{k|j} = \text{Tr}(\Pi^B_{k|j} \rho_j \Pi^B_{k|j})$ and
\begin{align}
	\rho_{j,k} &= \Pi_{k|j}^B \rho_j \Pi_{k|j}^B / p^B_{k|j},
	\\
	\rho_{\Pi^{AB}} &= \sum_{j,k} \Pi^{AB}_{jk} \rho_{j,k} \Pi^{AB}_{jk}.
\end{align}

 \noindent
\textbf{Entropy. }
 We denote by $S(\rho)$ the von Neumann \emph{entropy} of a quantum state $\rho$.
 The \emph{variation in entropy} induced by a measurement $\Pi^A$ is denoted by 
\begin{equation}
	\delta S_{\Pi^A} (\rho) = S_A (\rho_{\Pi^A}) - S_A (\rho).
\end{equation}
For a bipartite state $\rho$ the \emph{conditional entropy} of $B$ given $A$ is defined, respectively without and with measurement, by
\begin{align}
	&S_{B|A}(\rho) = S_{AB}(\rho) - S_{A}(\rho),
	\\
	&S_{B|\Pi^A} (\rho) = \sum_{j} p_{j}^A S_{AB} (\Pi_{j}^{A} \rho \Pi_{j}^{A}/p_{j}^A ).
\end{align}
To alleviate the notation, we sometimes write
\begin{align}
	S_{ \Pi^{A} |B}(\rho)  = S_{AB} (\rho_{\Pi^A}) - S_B( \rho_{\Pi^A}).
\end{align}
The \emph{variation in conditional entropy} after a measurement on $B$ conditioned on a preliminary measurement on $A$ is denoted by
\begin{align} 
	&\delta S_{B|\Pi^A} (\rho) = S_{B|A} (\rho_{\Pi^{AB}}) - S_{B|A} (\rho_{\Pi^A}).
	\label{eq-supp:deltaS-B-condPiA}
\end{align}
 
 \noindent
\textbf{Mutual information. } The \emph{mutual information} is defined, respectively without and with measurement, by
\begin{align}
 	&I_{A:B}(\rho) = S_B(\rho) - S_{B| A} (\rho),
	\\
	&J_{A:B}(\rho) =  S_B(\rho) - S_{B| \Pi^{A}} (\rho)  =  I_{A:B} (\rho_{\Pi^{A}}).
\end{align}
The \emph{conditional mutual information} is defined, without measurement, with measurement on $A$ and with measurement on $AB$ respectively, by
\begin{align} 
	&I_{A:B|C} (\rho)  = S_{A|C} (\rho)  - S_{A|BC} (\rho), 
	\\
	&J_{A:B|C} (\rho) =  I_{A:B|C} (\rho_{\Pi^{A}} ),
	\\
	& K_{A:B|C} (\rho) =  I_{A:B|C} (\rho_{\Pi^{AB}} ).
\end{align}
The \emph{tripartite mutual information} is defined, without measurement, with measurement on $A$ and with measurement on $AB$ respectively, by
\begin{align} 
	&I_{A:B:C} (\rho) = I_{A:C}(\rho) - I_{A:C|B}(\rho),
	\label{eq-supp:def-I-tripartite}
	\\
	&J_{A:B:C} (\rho) = I_{A:B:C} (\rho_{\Pi^{A}}),
	\label{eq-supp:def-J-tripartite}
	\\
	&K_{A:B:C} (\rho) = I_{A:B:C} (\rho_{\Pi^{AB}}).
	\label{eq-supp:def-K-tripartite}
\end{align}

 \noindent
\textbf{Discords. }The \emph{(bipartite) discord}, with a given projector $\Pi^A$, is
\begin{equation}
	d_{A;B}(\rho) =  S_{B|\Pi^A} (\rho) -S_{B|A} (\rho).
\end{equation}
For a tripartite state in which $A$ has already been measured, we use the notation
\begin{equation}
	d_{B;C} (\rho_{\Pi^A}) = S_{C|B} (\rho_{\Pi^{AB}}) -  S_{C|B} (\rho_{\Pi^A}) . 
	\label{eq-supp:d-BC-rhoPiA}
\end{equation}
We also use the notation
\begin{align}
	d_{B;\Pi^A C} ( \rho) & = S_{AC|B} (\rho_{\Pi^{AB}}) - S_{AC|B} (\rho_{\Pi^A}),
	\label{eq-supp:d-BPiAC-dd}
	\\
	d_{B; \Pi^A} ( \rho) & = S_{A|B}   (\rho_{\Pi^{AB}})  - S_{A|B}   (\rho_{\Pi^A}).
	\label{eq-supp:d-BPiA-S}
\end{align}
The \emph{(optimized bipartite) discord} is
\begin{equation}
	D_{A;B}(\rho) =  \min_{ \Pi^A } \left[S_{B|\Pi^A} (\rho) - S_{B|A} (\rho) \right].
\end{equation} 
The \emph{(optimized) tripartite discord} is
\begin{align}
	D_{A;B;C} (\rho) = & \min_{\Pi^{AB}  } \Big[  
 S_{B| \Pi^{A}} (\rho) + S_{C|  \Pi^{AB}} (\rho) - S_{BC|A} (\rho) \Big] 
\label{eq-supp:tripartitediscord}
\end{align}
where the min is over measurements of the form~\eqref{supp:twomeasurements}.
For given $\Pi^A$ and $\Pi^{AB}$, we call \emph{conditional mutual discords} and denote as follows the changes in conditional mutual information
\begin{align}
	\Delta_{B; \Pi^A |C} ( \rho) &=  J_{A:B|C} (\rho) - K_{A:B|C} (\rho)
	\label{eq-supp:Delta-BPiA-condC-JK}
	\\
	& = d_{B;\Pi^A C} ( \rho) -d_{B;C} (\rho_{\Pi^A}),
	\label{eq-supp:Delta-BPiA-condC-dd}
	\\
	\Delta_{A;B|C} (\rho)  
	& =  I_{A:B|C} (\rho) - J_{A:B|C} (\rho) 
	\\
	&= d_{A;BC} (\rho) - d_{A;C} (\rho),
	\label{eq-supp:Delta-AB-condC-dd}
	\\
	\Delta_{B;C|\Pi^A} (\rho)  
	& =  J_{B:C|A} (\rho) - K_{B:C|A} (\rho)  
	\\
	& = d_{B;\Pi^A C} (\rho) - d_{B;\Pi^A} (\rho),
	\label{eq-supp:Delta-BC-condPiA-dd}
\end{align}
and we also introduce the following notations for the change in tripartite information
\begin{align}
	\Delta_{A:B:C}(\rho)  
	&=  I_{A:B:C} (\rho) - J_{A:B:C} (\rho) 
	\label{eq-supp:def-Delta-tripartite}
	\\
	& = d_{A;B} (\rho) + d_{A;C} (\rho) - d_{A;BC} (\rho),
	\label{eq-supp:Delta-tripartite-ddd}
	\\
	\Delta_{B; \Pi^A ;C} ( \rho) &=  J_{A:B:C} (\rho) - K_{A:B:C} (\rho)
	\label{eq-supp:def-Delta-tripartite-PiA}
	\\
	&=d_{B;\Pi^A} (\rho) + d_{B;C} (\rho_{\Pi^A}) - d_{B; \Pi^A C} (\rho).
	\label{eq-supp:Delta-BPiAC-ddd}
\end{align}

\section{Conditional measurements}

In this section, we show that a measurement of the form of Eq. (3) in the main text acting on a zero discord state leaves it invariant. A zero discord state corresponds a state where there are zero quantum correlations, which is defined to be a state where the subsystem $ A $ has been measured.  Thus, assuming $d_{A;B}(\rho) = 0$, we would like to show that there are measurements $\Pi^{AB}$ of the form $\Pi^{AB}_{jk} = \Pi^A_j \otimes \Pi^B_{k|j}$ such that 
\begin{align}
\rho_{\Pi^A} = \sum_{jk}  \Pi_{jk}^{AB} \rho_{\Pi^A}  \Pi_{jk}^{AB} .
\label{aim1}
\end{align}
A general state after the measurement of subsystem $A$ can be written
\begin{align}
\rho_{\Pi^A} = \sum_{j} p_j^A |j\rangle \langle j |_A \otimes \rho^B_j
\label{rhopia}
\end{align}
where $ \rho^B_j $ is the state of subsystem $ B $ conditional on the $ j $th measurement outcome, and 
\begin{align}
\Pi_{j}^{A} = |j\rangle \langle j |_A 
\label{aprojector}
\end{align}
is the von Neumann projector.  The state $ \rho^B_j $ can always be diagonalized as 
\begin{align}
\rho^B_j = \sum_k \lambda_k^{(j)} | \lambda_k^{(j)}  \rangle \langle  \lambda_k^{(j)}  |_B, 
\end{align}
where $ \sum_k \lambda_k^{(j)} =1 $.  The state (\ref{rhopia}) can therefore be written 
\begin{align}
\rho_{\Pi^A} = \sum_{jk} p_j^A \lambda_k^{(j)} |j\rangle \langle j |_A \otimes | \lambda_k^{(j)}   \rangle
 \langle \lambda_k^{(j)} |_B .
\label{measuredstate2}
\end{align}
Taking $ \Pi_{k|j}^{B}  = \Pi_{k|j}^{B,\text{opt}} $ where 
\begin{align}
\Pi_{k|j}^{B,\text{opt}} = | \lambda_k^{(j)}  \rangle \langle  \lambda_k^{(j)}  | ,
\label{optimumbmeas}
\end{align}
one can verify that this satisfies (\ref{aim1}).

\section{Bipartite discord with two measurements}

In this section, we show that Eq. (4) in the main text gives the same result as Eq. (1) in the main text. Consider that the optimization of the measurement $ \Pi^{AB}   $ is done in two stages, where for a given 
$\Pi^A  $ measurement, the optimum (conditional) measurement $ \Pi^B $ is sought after.  The measured state is given by 
\begin{align}
\rho_{\Pi^{AB}} = \sum_{jk} p_j^A |j\rangle \langle j |_A \otimes \Pi^B_{k|j} \rho^B_j \Pi^B_{k|j}
\label{measuredstate1}
\end{align}
where we have performed a conditional measurement on (\ref{rhopia}), and  $ \Pi^B_{j,k}   $ do not necessarily consist of the eigenstates of  $ \rho^B_j $ as in the previous section. Write the general form of 
 $ \Pi^B_{k|j} $ as
\begin{align}
\Pi^B_{k|j} = | k; j \rangle \langle k;j  |_B .
\label{bprojector}
\end{align}
Then for any choice of $ \Pi^B_{k|j}  $ that do not necessarily involve the eigenstates of  $ \rho^B_j $, one has
\begin{align}
\rho_{\Pi^{AB}} = \sum_{jk} p_j^A  p_{k|j}^B |j\rangle \langle j |_A  \otimes 
| k; j \rangle \langle k;j  |_B
\label{measuredstate}
\end{align}
where 
\begin{align}
p_{k|j}^B = \sum_{k'} \lambda_{k'}^{(j)} |\langle k; j | \lambda_{k'}^{(j)}  \rangle|^2
\label{nonoptimalprobb}
\end{align}
is the conditional probability of the measurement result $ k $, for the outcome $ j $ on $ A $.  Finding the conditional entropy of (\ref{measuredstate}) evaluates to
\begin{align}
S_{B|A} (\rho_{\Pi^{AB}}) & = S_{AB} (\rho_{\Pi^{AB}}) - S_{A} (\rho_{\Pi^{AB}}) \nonumber \\
& = \sum_j p_j^A \log ( p_j^A )  - \sum_{jk} p_j^A  p_{k|j}^B \log ( p_j^A  p_{k|j}^B) \nonumber \\
& = - \sum_{jk} p_j^A  p_{k|j}^B \log ( p_{k|j}^B),
\end{align}
which takes the form of an average entropy of the distribution $ p_{k|j}^B $ over the outcomes $ j $.  
Eq. (\ref{nonoptimalprobb}) involves convolving two probability distributions, and the minimum entropy is reached with respect to the measurements $ \Pi^B_{k|j} $ when 
$  |\langle k; j | \lambda_{k'}^{(j)}  \rangle|^2 = \delta_{kk'} $.  Thus choosing  $ \Pi^B_{k|j} = \Pi_{k|j}^{B,\text{opt}}$ defined by Eq.~\eqref{optimumbmeas}, i.e., choosing it to coincide with the eigenstates of $  \rho^B_j $, minimizes $ S_{B|A} (\rho_{\Pi^{AB}}) $ for a given $ \Pi^A_j $.   We may thus equivalently write Eq. (4) as 
\begin{align}
 \min_{\Pi^{AB}}  & \left[ S_{B|A} ( \rho_{\Pi^{AB}} ) - S_{B|A}(\rho) \right]  \nonumber \\
& =  \min_{\Pi^{A}}  \left[ S_{B|A} ( \rho_{\Pi^{A} \otimes \Pi^{B,\text{opt}} } ) - S_{B|A}(\rho) \right] 
\label{equivalentdis}
\end{align}
From (\ref{measuredstate2}) it is evident that 
\begin{align}
\rho_{\Pi^{A} \otimes \Pi^{B,\text{opt}} }  = \rho_{\Pi^{A}}
\end{align}
hence 
\begin{align}
S_{B|A} ( \rho_{\Pi^{A} \otimes \Pi^{B,\text{opt}} } ) = S_{B|A} ( \rho_{\Pi^{A} } ) = S_{B| \Pi^A} (\rho) .
\end{align}
where we used Eq. (6) in the main text. Substituting this into (\ref{equivalentdis}) explicitly shows that it takes the same form as the definition of the discord, Eq. (1)  in the main text.

\section{Tripartite quantum discord}
\label{sec:tripartitediscord}

In this section, we show that the entropy of a measured tripartite system takes the form of Eq. (7) in the main text. After a measurement of the form of Eq. (3) in the main text, a general quantum state takes the form
\begin{align}
\rho_{\Pi^{AB}} = \sum_{jk}  p_j^A p_{k|j}^B |j\rangle \langle j |_A \otimes 
|k; j \rangle \langle k; j |_B \otimes \rho_{jk}^C
\end{align}
where we have taken the projectors of the form (\ref{aprojector}) and  (\ref{bprojector}).
The $  p_j^A $ and $ p_{k|j}^B  $ are the probabilities of the measurement outcomes. Diagonalizing the density matrix on $ C $, we can write
\begin{align}
\rho_{jk}^C = \sum_l \lambda_l^{(jk)} |\lambda_l^{(jk)}\rangle \langle \lambda_l^{(jk)} | . 
\end{align}
Evaluating the entropy of such as state yields
\begin{align}
S(\rho_{\Pi^{AB}}) = & \sum_j p_j^A \log p_j^A + \sum_{j} p_j^A \sum_{k} p_{k|j}^B \log p_{k|j}^B \nonumber \\
& + \sum_{jk} p_j^A p_{k|j}^B \sum_l \lambda_l^{(jk)} \log \lambda_l^{(jk)} .
\label{expandedrhopiab}
\end{align}
We can make the identifications
\begin{align}
S_A ( \rho_{\Pi^{AB}}) & = \sum_j p_j^A \log p_j^A  \\
S_{B | \Pi^A} ( \rho_{\Pi^{AB}}) & = \sum_{j} p_j^A \sum_{k} p_{k|j}^B \log p_{k|j}^B \\
S_{C| \Pi^{AB}} (  \rho) & = \sum_{jk} p_{jk}^{AB}  \sum_l\lambda_l^{(jk)} \log \lambda_l^{(jk)} ,
\end{align}
where we have used the notation $ p_{jk}^{AB} =  p_j^A p_{k|j}^B $ and the fact that 
\begin{align}
\rho_{\Pi^{AB}}^A & = \text{Tr}_{BC} (\rho_{\Pi^{AB}})  = \sum_j p_j^A  |j\rangle \langle j |_A \nonumber \\
\rho_{\Pi^{AB}}^{AB} & = \text{Tr}_C (\rho_{\Pi^{AB}})  = \sum_{jk}  p_j^A p_{k|j}^B |j\rangle \langle j |_A \otimes  |k; j \rangle \langle k; j |_B .
\end{align}
Eq. (7) in the main text then follows from (\ref{expandedrhopiab}).

\section{Properties of the tripartite discord}
\label{sec:prop-tri-discord}
\subsection{Non-negativity}
\label{sec:tri-disc-non-neg}
In this section, we show that the tripartite discord defined by Eq. (8) of the main 
text is non-negative.  

Using the equivalent form of Eq. (16) in the main text, and combining with Eq. (14) of the main text, we can equivalently write
\begin{align}
\label{eq:DABC-d-Delta}
D_{A;B;C} (\rho) = & \min_{\Pi^{AB}} \Big[ 
d_{A;BC} (\rho) +  \Delta_{B;C|\Pi^A} (\rho)  \Big] .
\end{align}
The quantity $ d_{A;BC} (\rho) $ is the bipartite discord without minimization between $ A$ and $ BC $ and is a non-negative quantity \cite{ollivier2001quantum}.  We show that the conditional discord $  \Delta_{B;C|\Pi^A} (\rho) $ is a non-negative quantity in (\ref{bcaconddis}) below.  It then follows that the tripartite discord is non-negative.

\subsection{$ \rho = \rho_{\Pi^{AB}} \implies D_{A;B;C}( \rho) = 0  $}

In this section, we show that a zero quantum correlations state implies a zero tripartite discord. 

For any state of the form $ \rho = \rho_{\Pi^{AB}}  $, from Eq. (7)
\begin{equation}
\begin{split}
&S_{ABC}(\rho_{\Pi^{AB}}) - S_A(\rho_{\Pi^{AB}}) - S_{B| \Pi^{A}} (\rho_{\Pi^{AB}})
\\
&= S_{C|  \Pi^{AB}} (\rho_{\Pi^{AB}} ) 
\end{split}
\end{equation}
where we used the fact that $ S_{C|  \Pi^{AB}} (\rho)  = S_{C|  \Pi^{AB}} (\rho_{\Pi^{AB}} ) $.  It then follows from the definition Eq. (8) and the definition of the conditional entropy $S_{BC|A}$ that $D_{A;B;C}( \rho) =0$.

\subsection{$ D_{A;B;C}( \rho) = 0  \implies    \rho = \rho_{\Pi^{AB}}  $}

In this section, we show that zero tripartite discord implies zero quantum correlations.  We generalize ideas that have been used in the case of the bipartite discord, see for example the PhD thesis of Datta~\cite{datta2008studies}.

Recall that the (optimized) tripartite discord is given by~\eqref{eq:DABC-d-Delta}. As noticed above in Sec. \ref{sec:tri-disc-non-neg}, each term inside the $\min$ is non-negative.

Then $D_{A;B;C} (\rho) = 0$ implies that there exists $\Pi^{AB}$ such that \textbf{(i)} $d_{A; BC} (\rho)=0$ and \textbf{(ii)} $\Delta_{B;C|\Pi^A}(\rho)=0$. (Recall that $d_{A; BC} (\rho)$ and $\Delta_{B;C|\Pi^A}$ depend respectively on $\Pi^A$ and $\Pi^{AB}$.) By a result of Proposition 1 in Ref. \cite{ollivier2001quantum}, point \textbf{(i)} implies that $\rho = \sum_j \Pi^A_j \rho \Pi^A_j = \sum_j p^A_j \rho_j$, where we recall that in our notations $p^A_j = \text{Tr}(\Pi^A_j \rho \Pi^A_j)$ and $\rho_j = \Pi^A_j \rho \Pi^A_j / p^A_j$. We want to show that point \textbf{(ii)} implies that for each $j$
\begin{equation}
\label{eq-supp:proof3-subgoal}
	\rho_j = \sum_k \Pi^B_{k|j} \rho_j \Pi^B_{k|j} = \sum_k p_{k|j}^{B} \rho_{j,k},
\end{equation}
where $p_{k|j}^{B} = \text{Tr}(\Pi^B_{k|j} \rho_j \Pi^B_{k|j})$ and $\rho_{j,k} = \Pi^B_{k|j} \rho_j \Pi^B_{k|j} / p_{k|j}^{B}$. This will yield the desired equality (namely, $\rho = \rho_{\Pi^{AB}}$) since we will then have
\begin{equation}
	\rho = \sum_{j,k} p^A_j p_{k|j}^{B} \rho_{j,k} = \rho_{\Pi^{AB}}.
\end{equation}
We split the proof into several steps. It will be useful to note that 
\begin{equation}
	\Delta_{B;C|\Pi^A}(\rho)=S_{C|AB}(\rho_{\Pi^{AB}}) - S_{C|AB}(\rho_{\Pi^{A}}).
	\label{eq-supp:proof3-Delta-S_CAB}
\end{equation}
\noindent
\textbf{Step 1:} We show that
\begin{align}
	&S_{C|AB}(\rho_{\Pi^{AB}}) - S_{C|AB}(\rho_{\Pi^{A}}) \nonumber
	\\
	&= \sum_{j}p^A_j \Big( S_{C|AB}(\rho_{j, \Pi^{B}_{.|j}}) - S_{C|AB}(\rho_j) \Big),
	\label{eq-supp:proof3-step1}
\end{align}
where $\rho_{j, \Pi^{B}_{.|j}} = \sum_k \Pi^B_{k|j} \rho_{j} \Pi^B_{k|j}= \sum_k p_{k|j}^{B} \rho_{j,k}$.

\begin{proof}
	First, since $\rho_{\Pi^A} = \sum_j p^A_j \rho_j$ is block diagonal, 
	\begin{align}
		S(\rho_{\Pi^A}) &= S_{ABC}(\rho_{\Pi^A}) 
		\\
		&= \sum_j p^A_j S(\rho_j) - \sum_j p^A_j\log(p^A_j)
		\\
		&= \sum_j p^A_j S(\rho_j) + S_A(\rho_{\Pi^A}),
	\end{align}
	and likewise,
	\begin{align}
		S_{AB}(\rho_{\Pi^A}) 
		&= \sum_j p^A_j S_{AB}(\rho_j) + S_A(\rho_{\Pi^A}).
	\end{align}
	We deduce that 
	\begin{equation}
	\label{eq-supp:proof3-step1-tmp1}
		S_{C|AB}(\rho_{\Pi^{A}}) = \sum_{j}p^A_j S_{C|AB}(\rho_j).
	\end{equation}
	Second, since $\rho_{\Pi^{AB}} = \sum_{j,k} p_{j,k}^{A,B} \rho_{j,k}$ is block diagonal,
	
	\begin{align}
		S(\rho_{\Pi^{AB}}) 
		= \sum_{j} p^A_j \sum_k p_{k|j}^{B} S(\rho_{j,k}) + S_{AB}(\rho_{\Pi^{AB}}).
		\label{eq-supp:proof3-1-tmp-0}
	\end{align}
	
	Moreover, 
	\begin{equation}
	\label{eq-supp:proof3-1-tmp}
		\sum_k p_{k|j}^{B} S(\rho_{j,k}) = S_{C|AB}(\rho_{j,\Pi^B_{.|j}}),
	\end{equation}
	which is proved as follows: since $\rho_{j,\Pi^B_{.|j}} = \sum_k p_{k|j}^{B}\rho_{j,k}$ is block diagonal, one can check that
	\begin{align}
		S(\rho_{j,\Pi^B_{.|j}}) &= S_{ABC}(\rho_{j,\Pi^B_{.|j}}) 
		\\
		&= \sum_{k} p_{k|j}^{B} S(\rho_{j,k}) + S_{AB}(\rho_{j,\Pi^B_{.|j}}).
		\label{eq-supp:proof3-1-SrhoPij}
	\end{align}
	
	We deduce from~\eqref{eq-supp:proof3-1-tmp-0} and~\eqref{eq-supp:proof3-1-tmp} that
	\begin{equation}
	\label{eq-supp:proof3-step1-tmp2}
		S_{C|AB}(\rho_{\Pi^{AB}}) = \sum_{j}p^A_j S_{C|AB}(\rho_{j, \Pi^{B}_{.|j}}).
	\end{equation}
	Combining~\eqref{eq-supp:proof3-step1-tmp1} and~\eqref{eq-supp:proof3-step1-tmp2} yields~\eqref{eq-supp:proof3-step1}.
\end{proof}

\noindent
\textbf{Step 2.} We next turn to the right hand side of~\eqref{eq-supp:proof3-step1} and show that
\begin{align}
\label{eq-supp:proof3-step2}
	S_{C|AB}(\rho_{j, \Pi^{B}_{.|j}}) - S_{C|AB}(\rho_j) 
	\geq 0.
\end{align}
\begin{proof}
	First, notice that we have~\eqref{eq-supp:proof3-1-tmp} and
	\begin{align}
		S_{C|AB}(\rho_j) &= S_{ABC}(\rho_{j}) - S_{AB}(\rho_{j}) .
	\end{align}
	Let us now introduce an extra subsystem, say $D$, which will contain a copy of $B$ in order to deal with the measurement. Let us define $\tilde \rho_j$ by
	\begin{align}
	\label{eq:proof-tripartite-tilde-rho-j}
		&\tilde \rho_j^{ABCD}
		\\
		&=
		\sum_{k,k'} \langle k;j | \rho_j | k';j \rangle \otimes | k;j \rangle\langle k';j |_B \otimes  | g_{k|j} \rangle\langle g_{k'|j} |_D
		\notag
	\end{align}
	where $| k;j \rangle$ is such that $\Pi^B_{k|j} = | k;j \rangle\langle k;j |$. Then, by strong subadditivity of von Neumann entropy, we obtain
	\begin{equation}
		S_{ABCD}(\tilde\rho_j) + S_{B}(\tilde\rho_j)
		\leq 
		S_{ABC}(\tilde\rho_j) + S_{BD}(\tilde\rho_j).
		\label{eq-supp:proof3-step2-subadd1}
	\end{equation}
	Note that, by construction of $\tilde \rho_j$,
	\begin{align}
		S_{ABCD}(\tilde\rho_j) &= S_{ABC}(\rho_j),
		\\
		S_{B}(\tilde\rho_j) &= H(\{p_{k|j}^{B}\}_k),
		\\
		S_{ABC}(\tilde\rho_j) &= S\left( \sum_{k} \langle k;j | \rho_j | k;j \rangle \otimes | k;j \rangle\langle k;j | \right) \nonumber
		\\
		&= H(\{p_{k|j}^{B}\}_k) + \sum_k p_{k|j}^{B} S_{AC}(\rho_{j,k}) \nonumber
		\\
		&= H(\{p_{k|j}^{B}\}_k) + \sum_k p_{k|j}^{B} S_{C}(\rho_{j,k}) ,
		\\
		S_{BD}(\tilde\rho_j) &= S_B(\rho_j) = S_{AB}(\rho_j),
	\end{align}
	where we used the fact that in $\rho_j$ the state of subsystem $A$ is fixed. Hence from~\eqref{eq-supp:proof3-step2-subadd1} we get
	\begin{equation}
		S_{ABC}(\rho_j)
		\leq 
		\sum_k p_{k|j}^{B} S_{C}(\rho_{j,k}) + S_{AB}(\rho_j),
	\end{equation}
	which amounts to~\eqref{eq-supp:proof3-step2}.
\end{proof}
By~\eqref{eq-supp:proof3-Delta-S_CAB}, ~\eqref{eq-supp:proof3-step1} and~\eqref{eq-supp:proof3-step2}, we see that $\Delta_{B;C|\Pi^A}(\rho)=0$ yields
\begin{equation}
\label{eq-supp:proof3-step2-consequence}
	S_{C|AB}(\rho_{j, \Pi^{B}_{.|j}}) - S_{C|AB}(\rho_j) 
	= 0
\end{equation}
for all $j$. We conclude the proof of~\eqref{eq-supp:proof3-subgoal} with the following step.

\noindent
\textbf{Step 3.} We show that
\begin{align}
\label{eq-supp:proof3-step3}
	\rho_j = \sum_k p_{k|j}^{B} \rho_{j,k}.
\end{align}
\begin{proof}
	Eq.~\eqref{eq-supp:proof3-step2-consequence} means that we have equality in~\eqref{eq-supp:proof3-step2-subadd1}. Using Theorem 6 in Ref. \cite{hayden2004structure}, there exists a decomposition of the Hilbert space of subsystem $B$ such that
	\begin{align}
		\tilde\rho_j = \bigoplus_{\alpha} q_{j,\alpha} \tilde \rho_{j,\alpha}^{ACB^L} \otimes \tilde\rho_{j,\alpha}^{DB^R},
	\end{align}
	where we recall that $\tilde \rho_j$ is defined by~\eqref{eq:proof-tripartite-tilde-rho-j}.
	Since $\tilde \rho_j$ is symmetric with respect to a measurement on $B$ or $D$, the decomposition must actually be of the following form
	\begin{align}
		\tilde\rho_j = \bigoplus_{\alpha} q_{j,\alpha} \tilde \rho_{j,\alpha}^{AC} \otimes \tilde\rho_{j,\alpha}^{DB}.
	\end{align}
	Let us consider the unitary $U$ such that
	\begin{equation}
		U | k;j \rangle_B \otimes | 0 \rangle_D
		= | k;j \rangle_B \otimes | g_{k|j} \rangle_D.
	\end{equation}
	We introduce the following notation for the diagonalization of $\rho_{j,\alpha}^B$:
	\begin{equation}
		\rho_{j,\alpha}^B = \sum_i \lambda_{j,\alpha,i} | \lambda_{j,\alpha,i} \rangle \langle \lambda_{j,\alpha,i}|. 
	\end{equation}
	Then,
	\begin{align}
		\rho_j^{ABC} 
		&= \langle 0_D | U^\dagger \tilde \rho_j^{ABCD} U | 0_D \rangle
		\\
		&= \sum_\alpha q_{j,\alpha} \rho_{j,\alpha}^{AC} \otimes \rho_{j,\alpha}^B
		\\
		&= \sum_{\alpha,i} q_{j,\alpha} \lambda_{j,\alpha,i} \rho_{j,\alpha}^{AC} \otimes | \lambda_{j,\alpha,i} \rangle \langle \lambda_{j,\alpha,i}|.
	\end{align}
	Let us write the diagonalization of $\rho_{j,\alpha}^{AC}$ as:
	\begin{equation}
		\rho_{j,\alpha}^{AC} = |j \rangle \langle j |_A \otimes \sum_\ell \gamma_{j,\alpha,\ell} |\gamma_{j,\alpha,\ell}\rangle\langle \gamma_{j,\alpha,\ell}|_C. 
	\end{equation}
	Then, denoting $\delta_{j,\alpha,i,\ell} = q_{j,\alpha} \lambda_{j,\alpha,i} \gamma_{j,\alpha,\ell}$, we have
	\begin{align}
		\rho_j^{ABC} 
		&= \sum_{\alpha,i,\ell} \delta_{j,\alpha,i,\ell}  |j , \gamma_{j,\alpha,\ell}, \lambda_{j,\alpha,i} \rangle \langle j , \gamma_{j,\alpha,\ell}, \lambda_{j,\alpha,i}  |.
	\end{align}
	
	We obtain~\eqref{eq-supp:proof3-step3} by a suitable relabeling: if replace each pair $(\alpha,i)$ by some index $k$, then we have $\Pi^B_{k|j} = | \lambda_{j,\alpha,i} \rangle \langle \lambda_{j,\alpha,i}  |_B$ and $p^B_{k|j} = \text{Tr}(\Pi^B_{k|j} \rho_j \Pi^B_{k|j}) = q_{j,\alpha} \lambda_{j,\alpha,i}$.
\end{proof}

\subsection{Reduction of tripartite discord to bipartite discord}

In this section, we show that the tripartite discord reduces to the standard bipartite discord for bipartite correlated states.

\subsubsection{AB correlated states}

The definition of the tripartite discord as given in Eq. (8) of the main text contains three terms to be evaluated.  The first term can be decomposed as
\begin{equation}
	S_{BC|A}(\rho) = S_{ABC}(\rho) - S_{A}(\rho).
\end{equation}
The right hand side of Eq. (8) can then be written as
\begin{equation}
\label{eq-supp:eq8-rhs-4terms}
	 \min_{\Pi^{AB}  } \Big[  - S_{ABC}(\rho) + S_{A}(\rho)
 + S_{B| \Pi^{A}} (\rho) + S_{C|  \Pi^{AB}} (\rho) \Big] 
\end{equation}
Substituting $ \rho = \rho^{AB} \otimes \rho^C $ into these terms we obtain
\begin{align}
S_{ABC} ( \rho^{AB} \otimes \rho^C) & = S_{AB} (\rho^{AB}) + S_C (\rho^C)  \\
S_A ( \rho^{AB} \otimes \rho^C)  & = S_A (  \rho^{AB} ) \\
S_{B|\Pi^A} (\rho^{AB} \otimes \rho^C) & = S_{B|\Pi^A} (\rho^{AB} )  \\
S_{C|  \Pi^{AB}} (\rho^{AB} \otimes \rho^C) & = S_{C|AB} (\rho^{AB}_{\Pi^{AB}} \otimes \rho^C) \nonumber \\
& = S_{ABC} (\rho^{AB}_{\Pi^{AB}} \otimes \rho^C) - S_{AB} (\rho^{AB}_{\Pi^{AB}}) \nonumber \\
& = S_C (  \rho^C) .
\end{align}
The tripartite discord then reduces to 
\begin{align}
D_{A;B;C} & (\rho_{AB} \otimes \rho_C) \nonumber \\
& = \min_{\Pi^{AB}  } \Big[ 
	-S_{AB} (\rho^{AB}) + S_A (  \rho^{AB} ) + S_{B|\Pi^A} (\rho^{AB} ) \Big]  \nonumber \\
& = \min_{\Pi^{A}  } \Big[ 
	-S_{B|A} (\rho^{AB})  + S_{B|\Pi^A} (\rho^{AB} ) \Big] \nonumber \\
& = D_{A;B} (\rho_{AB} )
\end{align}
where we used the definition Eq. (1) of the main text.

\subsubsection{BC correlated states}

Evaluating the four terms in Eq.~\eqref{eq-supp:eq8-rhs-4terms} with  $ \rho = \rho^{BC} \otimes \rho^A $ we obtain
\begin{align}
S_{ABC} (  \rho^{BC} \otimes \rho^A) & = S_{BC} (\rho^{BC}) + S_A (\rho^A)  \\
S_A (  \rho^{BC} \otimes \rho^A)  & = S_A (  \rho^{A} ) \\
S_{B|\Pi^A} ( \rho^{BC} \otimes \rho^A) & = S_{B|A} ( \rho^{A}_{\Pi^A} \otimes \rho^B ) = S_B(\rho^B )  .
\end{align}
For the last term, we have
\begin{align}
S_{C|  \Pi^{AB}} & ( \rho^{BC} \otimes \rho^A)  \nonumber  \\
& = 
\sum_{jk} p_{jk}^{AB} S_{ABC} ( \Pi_j^A \rho^A \Pi_j^A \otimes \Pi_{k|j}^B \rho^{BC} \Pi_{k|j}^B / p_{jk}^{AB} ) \nonumber \\
& = \sum_{jk} p_j^A p_{k|j}^{B} S_{ABC} ( |j\rangle \langle j |_A \otimes \Pi_{k|j}^B \rho^{BC} \Pi_{k|j}^B / p_{k|j}^{B})  \nonumber \\
& = \sum_{jk} p_j^A p_{k|j}^{B} S_{BC} (\Pi_{k|j}^B \rho^{BC} \Pi_{k|j}^B / p_{k|j}^{B}).
\end{align}
The tripartite discord then reduces to 
\begin{align}
D_{A;B;C}&  ( \rho^{BC} \otimes \rho^A) = \min_{\Pi^{AB}  } \Big[ 
	-S_{C|B} (\rho^{BC})  \nonumber \\
	& + \sum_{jk} p_j^A p_{k|j}^{B} S_{BC} (\Pi_{k|j}^B \rho^{BC} \Pi_{k|j}^B / p_{k|j}^{B})   \Big] .
\end{align}
The state $ \rho^{BC} $ has no dependence on the measurement outcome $ j $, hence the optimal operators on $ B $ are independent of $ j $, giving $ \Pi_{k|j}^B = \Pi_{k}^B $.  The expression can thus equivalently be written
\begin{align}
D_{A;B;C}&  ( \rho^{BC} \otimes \rho^A) = \min_{\Pi^{AB}  } \Big[ 
	-S_{C|B} (\rho^{BC})  \nonumber \\
	& + \sum_{k} p_{k}^{B} S_{BC} (\Pi_{k}^B \rho^{BC} \Pi_{k}^B / p_{k}^{B})   \Big] \nonumber \\
& =  \min_{\Pi^{B}  } \Big[ - S_{C|B} (\rho^{BC}) + S_{C|\Pi^B} (\rho^{BC}) \Big] \nonumber \\
& = D_{B;C} (\rho^{BC}) 
\end{align}
where we used the fact that $ \sum_{j} p_j^A  = 1 $.

\subsubsection{AC correlated states}

Evaluating the four terms in Eq.~\eqref{eq-supp:eq8-rhs-4terms}  with  $ \rho = \rho^{AC} \otimes \rho^B $ we obtain
\begin{align}
S_{ABC} (\rho^{AC} \otimes \rho^B ) & = S_{AC} (\rho^{AC}) + S_B (\rho^B)  \\
S_A (\rho^{AC} \otimes \rho^B )  & = S_A (  \rho^{AC} ) \\
S_{B|\Pi^A} (\rho^{AC} \otimes \rho^B ) & = S_{B|A} ( \rho^{A}_{\Pi^A} \otimes \rho^B ) = S_B(\rho^B ).
\end{align}
For the last term, we have
\begin{align}
S_{C|  \Pi^{AB}} & (\rho^{AC} \otimes \rho^B )  \nonumber  \\
& = 
\sum_{jk} p_{jk}^{AB} S_{ABC} ( \Pi_j^A \rho^{AC} \Pi_j^A \otimes \Pi_{k|j}^B \rho^{B} \Pi_{k|j}^B / p_{jk}^{AB} ) \nonumber \\
& = \sum_{jk} p_j^A p_{k|j}^{B} S_{ABC} (\Pi_j^A \rho^{AC} \Pi_j^A  \otimes |k;j \rangle \langle k;j |_B/ p_j^A ) \nonumber \\
& = \sum_{jk} p_j^A p_{k|j}^{B} S_{AC} ( \Pi_j^A \rho^{AC} \Pi_j^A/ p_j^A   ) \nonumber \\
& = \sum_{j} p_j^A S_{AC} (\Pi_j^A \rho^{AC} \Pi_j^A/ p_j^A ) = S_{C|\Pi^A} ( \rho^{AC}).
\end{align}
The tripartite discord then reduces to 
\begin{align}
D_{A;B;C} & (\rho^{AC} \otimes \rho^B) \nonumber \\
	& = \min_{\Pi^{AB} } \Big[ 
		-S_{C|A} (\rho^{AC}) + S_{C|\Pi^A}  (\rho^{AC} ) \Big] \nonumber \\
	& = D_{A;C} (\rho_{AC} )
\end{align}
where the optimization over $ \Pi^{AB} $ can be reduced to $ \Pi^{A} $ since there is no dependence on subsystem $ B $ of the function.

\section{Multipartite quantum discord}

In this section, we show how we obtain Eq. (9) in the main text. The same steps are followed as in Sec. \ref{sec:tripartitediscord}.  The state of an $ N $-partite state after $ N -1 $ conditional measurements is
\begin{align}
 &\rho_{ \Pi^{A_1 \dots A_{N-1} }}  \nonumber \\
 & = \sum_{j_1 \dots j_{N}} p_{j_1}^{A_1} p_{j_2 | j_1 }^{A_2} \dots p_{j_{N-1} | j_1 \dots j_{N-2} }^{A_{N-1}} \lambda_{j_N}^{(j_1 \dots j_{N-1})} \nonumber \\
&  |j_1 \rangle \langle j_1 |_{A_1} \otimes |j_2; j_1 \rangle \langle j_2 ;j_1 |_{A_2} \otimes \dots \nonumber \\
& \otimes |j_{N-1}; j_1 \dots j_{N-2} \rangle \langle j_{N-1}; j_1 \dots j_{N-2}  |_{A_{N-1}} \nonumber \\
& \otimes | \lambda_{j_N}^{(j_1 \dots j_{N-1})} \rangle \langle \lambda_{j_N}^{(j_1 \dots j_{N-1})} |_{A_N} .
\end{align}
Evaluating the entropy of this state gives
\begin{align}
&S( \rho_{ \Pi^{A_1 \dots A_{N-1} }} )  = \sum_{j_1} h(p_{j_1}^{A_1}) \nonumber \\
&  + \sum_{j_1 } p_{j_1}^{A_1} \sum_{j_2 } p_{j_2 | j_1 }^{A_2} \log p_{j_2 | j_1 }^{A_2}  + \dots \nonumber \\
& + \sum_{j_1 \dots j_{N-1}} p_{j_1}^{A_1} p_{j_2 | j_1 }^{A_2} \dots p_{j_{N-1} | j_1 \dots j_{N-2} }^{A_{N-1}} 
\sum_{j_N } h(\lambda_{j_N}^{(j_1 \dots j_{N-1})}) 
\end{align}
where for the sake of brevity we used the notation $h(x) = x \log x$.
Making similar associations as in Sec. \ref{sec:tripartitediscord} we obtain the equality for the measured states
\begin{align}
& S_{A_1 \dots A_N}(\rho_{ \Pi^{A_1 \dots A_{N-1} }} )=
S_{A_1} (\rho_{ \Pi^{A_1 \dots A_{N-1} }}  )   \nonumber \\
& +  S_{A_2| \Pi^{A_1} } (\rho_{ \Pi^{A_1 \dots A_{N-1} }}  ) + \dots + S_{A_N|  \Pi^{A_1 \dots A_{N-1} } }  (\rho ) .  
\end{align}
The difference between the left and right sides is identified as the multipartite discord.

\section{Entropy flux for bipartite systems}

In this section, we show the entropy flux for the various contributions of entropy in a bipartite system as shown in Fig. 2(c) of the main text.

\subsection{Conditional entropy $ S_{A|B} $}

The conditional entropy in subsystem $ A $ prior to performing a  measurement $ \Pi^A $ is given by
\begin{align}
S_{A|B} (\rho) & = S_{AB} (\rho) - S_{B} (\rho)  \nonumber \\
& = S_{B|A} (\rho) + S_{A} (\rho) - S_{B} (\rho) .  
\end{align}
After the measurement, the entropy is 
\begin{align}
 	S_{ \Pi^{A} |B}(\rho) & \equiv S_{AB} (\rho_{\Pi^A}) - S_B( \rho_{\Pi^A})
	\nonumber \\
	& = S_{B| \Pi^{A}}(\rho) + S_A (\rho_{\Pi^A}) - S_B (\rho) ,
\end{align}
where we used Eq. (6) in the main text and the fact that $ \rho^B = \text{Tr}_A \rho = \text{Tr}_A \rho_{\Pi^A} $. 

The change in the conditional entropy is thus
\begin{align}
\delta S_{ \Pi^{A} |B}(\rho) & =  	S_{ \Pi^{A} |B}(\rho) - S_{A|B} (\rho)  \nonumber \\
& = S_{B| \Pi^{A}}(\rho) - S_{B|A} (\rho) + S_A (\rho_{\Pi^A}) - S_{A} (\rho) \nonumber \\
& = d_{A;B} ( \rho) + \delta S_{\Pi^A} (\rho), 
\end{align}
where the biparite discord without minimization is defined as 
\begin{align}
d_{A;B} ( \rho) & =S_{B| \Pi^{A}}(\rho) - S_{B|A} (\rho) 
\label{eq-supp:bipartite-d-no-min}
\end{align}
and the entropy change in subsystem $ A $ is
\begin{align}
\delta S_{\Pi^A} (\rho)& =  S_A (\rho_{\Pi^A}) - S_{A} (\rho) .
\end{align}

\subsection{Mutual entropy $ I_{A:B} $}

The mutual information  prior to performing a  measurement $ \Pi^A $ is given by
\begin{align}
 I_{A:B} (\rho) & = S_{A} (\rho)  + S_{B} (\rho)  - S_{AB} (\rho) \nonumber \\
& = S_{B} (\rho) - S_{B|A} (\rho) .  
\end{align}
After the measurement, the mutual information is
\begin{align}
 J_{A:B} (\rho) & = S_{B} (\rho)  - S_{B|\Pi^A} (\rho) \nonumber \\
& = I_{A:B} ( \rho_{\Pi^A} )  .
\end{align}
The change in the mutual information is thus
\begin{align}
\delta  J_{A:B} & =   J_{A:B} (\rho)  -  I_{A:B} (\rho)  \nonumber \\
& =S_{B|A} (\rho) - S_{B|\Pi^A} (\rho) \nonumber \\
& = -d_{A;B} ( \rho)  .
\end{align}

\subsection{Conditional entropy $ S_{B|A} $}

The conditional entropy in subsystem $ B $ prior to performing a  measurement $ \Pi^A $ is given by
%
$S_{B|A} (\rho).$ 
%
After the measurement, the entropy is 
%
$S_{B| \Pi^{A}}(\rho).$ 
%
By Eq.~\eqref{eq-supp:bipartite-d-no-min}, the change in the conditional entropy is thus
\begin{align}
\delta S_{B| \Pi^{A}}(\rho) & =  S_{B| \Pi^{A}}(\rho) - S_{B|A} (\rho)  \nonumber \\
& = d_{A;B} ( \rho) . 
\end{align}

\section{Entropy flux for tripartite systems}

In this section, we show the entropy flux for the various contributions of entropy in a tripartite system as shown in Fig. 2(e) and 2(f) of the main text.

\subsection{Conditional entropy $ S_{A|BC} $}

The conditional entropy in subsystem $ A $ with no measurements is
\begin{align}
S_{A|BC} (\rho) & = S_{ABC} (\rho) - S_{BC} (\rho)  \nonumber \\
& =  S_{BC|A} (\rho) + S_A  (\rho) - S_{BC} (\rho)  \nonumber \\
& =  S_{C|AB} (\rho) + S_{A|B}  (\rho) - S_{C|B} (\rho) \nonumber \\
& =  S_{AC|B} (\rho) - S_{C|B} (\rho) . \label{eq-supp:condSA-no-meas}
\end{align}
After one measurement $ \Pi^A $, the conditional entropy is
\begin{align}
S_{A|BC} (\rho_{\Pi^A}) & = S_{BC|\Pi^A} (\rho) + S_{A}(\rho_{\Pi^A}) - S_{BC} (\rho) \nonumber \\
& = S_{C|AB} (\rho_{\Pi^A}) + S_{A|B}  (\rho_{\Pi^A}) - S_{C|B} (\rho_{\Pi^A}) . \label{eq-supp:condSA-1-meas}
\end{align}
After two measurements $ \Pi^{AB} $,  the conditional entropy is
\begin{align}
S_{A|BC} (\rho_{\Pi^{AB}}) 
	= & S_{C|\Pi^{AB}} (\rho) + S_{A|B}(\rho_{\Pi^{AB}}) - S_{C|B} (\rho_{\Pi^{AB}}) \nonumber \\
	= & S_{AC|B} (\rho_{\Pi^{A}}) - S_{C|B} (\rho_{\Pi^{A}}). \label{eq-supp:condSA-2-meas}
\end{align}

The change after the first measurement is, by Eq.~\eqref{eq-supp:condSA-no-meas} and Eq.~\eqref{eq-supp:condSA-1-meas},
\begin{align}
S_{A|BC} (\rho_{\Pi^A}) - S_{A|BC} (\rho) & = d_{A;BC} (\rho) + \delta S_{\Pi^A} (\rho) .
\end{align}
The change after the second measurement is, by Eq.~\eqref{eq-supp:condSA-no-meas} and Eq.~\eqref{eq-supp:condSA-2-meas},
\begin{align}
S_{A|BC} (\rho_{\Pi^{AB}}) - S_{A|BC} (\rho_{\Pi^A}) = \Delta_{B; \Pi^A |C} ( \rho) ,
\end{align}
where we used~\eqref{eq-supp:Delta-BPiA-condC-dd}, \eqref{eq-supp:d-BPiAC-dd} and~\eqref{eq-supp:d-BC-rhoPiA}. 

\subsection{Conditional entropy $ S_{B|AC} $}

The conditional entropy in subsystem $ B $ with no measurements is
\begin{align}
S_{B|AC} (\rho) & = S_{ABC} (\rho) - S_{AC} (\rho)  \nonumber \\
& =  S_{BC|A} (\rho)  - S_{C|A} (\rho)  \nonumber \\
& =  S_{C|AB} (\rho) + S_{B|A}  (\rho) - S_{C|A} (\rho).  
\end{align}
After one measurement $ \Pi^A $, it is
\begin{align}
S_{B|AC} (\rho_{\Pi^A}) & = S_{BC|\Pi^A} (\rho) - S_{C|\Pi^A} (\rho) \\
& =  S_{C|AB} (\rho_{\Pi^A}) + S_{B|\Pi^A}  (\rho) - S_{C|\Pi^A} (\rho) . 
\end{align}
After two measurements $ \Pi^{AB} $, it is
\begin{align}
S_{B|AC} (\rho_{\Pi^{AB}})  = &  S_{C|\Pi^{AB}} (\rho) + S_{B|A}(\rho_{\Pi^{AB}}) - S_{C|A} (\rho_{\Pi^{AB}}) .
\end{align}

The change after the first measurement is
\begin{align}
S_{B|AC} (\rho_{\Pi^A}) - S_{B|AC} (\rho) & = \Delta_{A;B|C} (\rho),
\end{align}
where we used~\eqref{eq-supp:Delta-AB-condC-dd}.
The change after the second measurement is
\begin{align}
& S_{B|AC} (\rho_{\Pi^{AB}})  - S_{B|AC} (\rho_{\Pi^A})  \nonumber \\
& = \Delta_{B;C| \Pi^A}  (\rho)  + \delta S_{B| \Pi^A} ( \rho) 
\end{align}
where we used~\eqref{eq-supp:Delta-BC-condPiA-dd}, \eqref{eq-supp:d-BPiA-S} and~\eqref{eq-supp:deltaS-B-condPiA}
We note that we can also write
\begin{align}
 \Delta_{B;C| \Pi^A}  (\rho) = &  S_{AC|B} (\rho_{\Pi^{AB}})  - 
S_{A|B}   (\rho_{\Pi^{AB}})   \nonumber \\
& - S_{AC|B} (\rho_{\Pi^A}) +S_{A|B}   (\rho_{\Pi^A}) \nonumber \\
= &  S_{C|AB} (\rho_{\Pi^{AB}}) - S_{C|AB} (\rho_{\Pi^A}).
\label{somerel}
\end{align}

\subsection{Conditional entropy $ S_{C|AB} $}

The conditional entropy in subsystem $ C $ with no measurements is
\begin{align}
S_{C|AB} (\rho) & = S_{ABC} (\rho) - S_{AB} (\rho)  \nonumber \\
& =  S_{BC|A} (\rho)  - S_{B|A} (\rho)  .
\end{align}
After one measurement $ \Pi^A $, it is
\begin{align}
S_{C|AB} (\rho_{\Pi^A}) = S_{BC|\Pi^A} (\rho) - S_{B|\Pi^A} (\rho).  
\end{align}
After two measurements $ \Pi^{AB} $, it is
\begin{align}
& S_{C|AB} (\rho_{\Pi^{AB}})  = S_{C|\Pi^{AB}} (\rho).
\end{align}

The change after the first measurement is
\begin{align}
S_{C|AB} (\rho_{\Pi^A}) - S_{C|AB} (\rho) & = \Delta_{A;C|B} (\rho),
\end{align}
where we used~\ref{eq-supp:Delta-AB-condC-dd}.
The change after the second measurement is
\begin{align}
S_{C|AB} (\rho_{\Pi^{AB}})  -S_{C|AB} (\rho_{\Pi^A})  = \Delta_{B;C|\Pi^A} (\rho), 
\end{align}
which follows from (\ref{somerel}).

\subsection{Conditional mutual information $ I_{A:B|C} $}

The conditional mutual information between subsystems $ AB $ with no measurements is
\begin{align}
I_{A:B|C} (\rho) & = S_{AC} (\rho) + S_{BC} (\rho) - S_{ABC} (\rho) -S_{C} (\rho) \nonumber \\
& =  S_{C|A} (\rho)  - S_{BC|A} (\rho) + S_{BC} (\rho)- S_{C} (\rho) \nonumber \\
& = S_{C|A} (\rho) + S_{C|B} (\rho) -  S_{C|AB} (\rho) + S_A (\rho) \nonumber \\
&  - S_{A|B} (\rho)- S_{C} (\rho) .
\end{align}
After one measurement $ \Pi^A $, it is
\begin{align}
I_{A:B|C} (\rho_{\Pi^A}) & = S_{C|\Pi^A} (\rho) - S_{BC|\Pi^A} (\rho)   \nonumber \\
&\qquad + S_{BC} (\rho)- S_{C} (\rho)
 \nonumber \\
& =  S_{C|\Pi^A} (\rho) + S_{C|B} (\rho) -  S_{C|AB}  (\rho_{\Pi^A}) \nonumber \\
&  \qquad  + S_A  (\rho_{\Pi^A}) - S_{A|B} (\rho_{\Pi^A}) - S_{C} (\rho) .
\end{align}
After two measurements $ \Pi^{AB} $, it is
\begin{align}
 &  I_{A:B|C} (\rho_{\Pi^{AB}})= S_{C|\Pi^A} (\rho_{\Pi^{AB}}) +  S_{C|\Pi^B} (\rho_{\Pi^{AB}}) \nonumber \\
& - S_{C|\Pi^{AB}} (\rho) + S_A (\rho_{\Pi^{AB}})  - S_{A|B} (\rho_{\Pi^{AB}})- S_{C} (\rho) .
\end{align}

The change after the first measurement is
\begin{align}
I_{A:B|C} (\rho_{\Pi^A}) - I_{A:B|C} (\rho) & = d_{A;C} ( \rho) - d_{A;BC} ( \rho) \nonumber \\
& = - \Delta_{A;B|C} (\rho). 
\end{align}
The change after the second measurement is
\begin{align}
I_{A:B|C} (\rho_{\Pi^{AB}}) - I_{A:B|C} (\rho) & = - \Delta_{B; \Pi^A |C} ( \rho)
\end{align}
where we used the fact that $ S_{C|A} (\rho_{\Pi^{AB}}) = S_{C|\Pi^{A}} (\rho)  $ and $S_{A}(\rho_{\Pi^{AB}}) = S_{A}(\rho_{\Pi^{A}})$.

\subsection{Conditional mutual information $ I_{B:C|A} $}

The conditional mutual information between subsystems $ BC $ with no measurements is
\begin{align}
I_{B:C|A} (\rho) & = S_{AB} (\rho) + S_{AC} (\rho) - S_{ABC} (\rho) -S_{A} (\rho) \nonumber \\
& =  S_{B|A} (\rho)  + S_{C|A} (\rho) - S_{BC|A} (\rho)  \nonumber \\
& = S_{C|A} (\rho)  - S_{C|AB} (\rho) .
\end{align}
After one measurement $ \Pi^A $, it is
\begin{align}
I_{B:C|A} (\rho_{\Pi^A}) & = S_{B|\Pi^A} (\rho)  + S_{C|\Pi^A} (\rho) - S_{BC|\Pi^A} (\rho) \nonumber \\
 & =    S_{C|\Pi^A} (\rho)  - S_{C|AB} (\rho_{\Pi^A}) .
\end{align}
After two measurements $ \Pi^{AB} $, it is
\begin{align}
 I_{B:C|A} (\rho_{\Pi^{AB}})= S_{C|\Pi^A} (\rho_{\Pi^{AB}}) -  S_{C|\Pi^{AB}} (\rho) .
\end{align}

The change after the first measurement is
\begin{align}
 I_{B:C|A} (\rho_{\Pi^A}) - I_{A:B|C} (\rho) & = \Delta_{A:B:C} (\rho) ,
\end{align}
where we used~\eqref{eq-supp:Delta-tripartite-ddd}.
The change after the second measurement is
\begin{align}
 I_{B:C|A} (\rho_{\Pi^{AB}}) - I_{B:C|A} (\rho_{\Pi^A}) & = 
- \Delta_{B; C | \Pi^A } ( \rho) 
\end{align}
since $ S_{C|A} (\rho_{\Pi^{AB}}) = S_{C|\Pi^{A}} (\rho) $.

\subsection{Conditional mutual information $ I_{A:C|B} $}

The conditional mutual information between subsystems $ AC $ with no measurements is
\begin{align}
I_{A:C|B} (\rho) & = S_{AB} (\rho) + S_{BC} (\rho) - S_{ABC} (\rho) -S_{B} (\rho) \nonumber \\
	& = S_{A|B}(\rho) + S_{C|B}(\rho) - S_{AC|B}(\rho) .
\end{align}
After one measurement $ \Pi^A $, it is
\begin{align}
I_{A:C|B} (\rho_{\Pi^A}) & = S_{B|\Pi^A} (\rho)  + S_{C|B} (\rho) - S_{BC|\Pi^A} (\rho) \nonumber \\
	& = S_{A|B}(\rho_{\Pi^A})+ S_{C|B}(\rho_{\Pi^A}) - S_{AC|B}(\rho_{\Pi^A}) .
\end{align}
After two measurements $ \Pi^{AB} $, it is
\begin{align}
&I_{A:C|B} (\rho_{\Pi^{AB}}) \nonumber \\
	& = S_{A|B}(\rho_{\Pi^{AB}}) + S_{C|B}(\rho_{\Pi^{AB}}) - S_{AC|B}(\rho_{\Pi^{AB}}) .
\end{align}

The change after the first measurement is
\begin{align}
I_{A:C|B} (\rho_{\Pi^A}) - I_{A:C|B} (\rho) & = - \Delta_{A;C|B} (\rho),
\end{align}
where we used~\eqref{eq-supp:Delta-AB-condC-dd}.
The change after the second measurement is
\begin{align}
I_{A:C|B} (\rho_{\Pi^{AB}}) -I_{A:C|B} (\rho_{\Pi^A}) & = 
\Delta_{B:\Pi^A:C} (\rho), 
\end{align}
where we used~\eqref{eq-supp:Delta-BPiAC-ddd}.

\subsection{Tripartite mutual information $ I_{A:B:C} $}

By~\eqref{eq-supp:def-J-tripartite} and \eqref{eq-supp:def-Delta-tripartite}, we see that the change after the first measurement, $\Pi^A$, is
\begin{align}
I_{A:B:C} (\rho_{\Pi^A}) -I_{A:B:C} (\rho) & = - \Delta_{A:B:C} (\rho). 
\end{align}
By~\eqref{eq-supp:def-J-tripartite}, \eqref{eq-supp:def-K-tripartite} and \eqref{eq-supp:def-Delta-tripartite-PiA}, we see that the change after two measurements, $\Pi^{AB}$, is
\begin{align}
I_{A:B:C} (\rho_{\Pi^{AB}}) - I_{A:B:C} (\rho_{\Pi^A}) & = 
- \Delta_{B:\Pi^A:C} (\rho).   
\end{align}

\section{Properties of the conditional discord $ \Delta_{A;B|C} $}
\label{sec-supp:prop-condi-discord}

In this section, we describe some properties of the conditional discord  $ \Delta_{A;B|C} $, as defined in Eq. (12) of the main text.

\subsection{Non-negativity}

Here we show that for any state $\rho$ the conditional discords  $ \Delta_{A;B|C}( \rho) \geq 0$ and $ \Delta_{B;C|\Pi^A}( \rho) \geq 0$.  

For the first conditional discord, from the definition we write
\begin{align}
\Delta_{A;B|C} (\rho)  =  I_{A:B|C} (\rho) -  I_{A:B|C} (\rho_{\Pi^{A}} )  .
\end{align}
The conditional mutual information can be written as
\begin{align}
I_{A:B|C}(\rho) = S_{B|C}(\rho) + S_{C|A}(\rho) - S_{BC|A}(\rho) .  
          \label{eq-supp:defQCMIforJ}
\end{align}
We therefore have
\begin{align}
\Delta_{A;B|C} (\rho) & = \left[ S_{BC|A} (\rho_{\Pi^A})  - S_{C|A} (\rho_{\Pi^A}) \right]  \nonumber \\
& - \left[ S_{BC|A}(\rho) - S_{C|A}(\rho) \right] , 
\end{align}
which is non-negative by concavity of the function $\rho \mapsto S_{BC|A} (\rho) - S_{C|A} (\rho) = S_{B|CA}(\rho) $ (see e.g.~\cite[Exercise 11.7.5 page 320]{MR3645110}).

For the second conditional discord, from the definition we write
\begin{align}
\Delta_{B;C|\Pi^A} (\rho)  =  I_{B:C|A} (\rho_{\Pi^{A}}) -  I_{B:C|A} (\rho_{\Pi^{AB}})  .
\end{align}
The conditional mutual information can be written as
\begin{align}
I_{B:C|A} (\rho) =  S_{B|A}(\rho) + S_{C|A}(\rho)- S_{BC|A}(\rho) .  
          \label{eq-supp:defQCMIforJ2}
\end{align}
We therefore have
\begin{align}
\Delta_{B;C|\Pi^A} (\rho) & = \left[ S_{BC|A} (\rho_{\Pi^{AB}})  -  S_{B|A} (\rho_{\Pi^{AB}}) \right]  \nonumber \\
& - \left[ S_{BC|A}(\rho_{\Pi^{A}}) - S_{B|A} (\rho_{\Pi^{A}}) \right] , 
\label{bcaconddis}
\end{align}
where we used the fact that $ S_{C|A} ( \rho_{\Pi^{AB}}) = S_{C|A} ( \rho_{\Pi^{A}}) $.  The above is non-negative by concavity of the function $\rho \mapsto S_{BC|A} (\rho) - S_{B|A} (\rho) $.

\subsection{Reduction to bipartite discord}

Here we show that for the state $\rho = \rho^{AB} \otimes \rho^C$, the conditional discord reduces to the bipartite discord, up to the basis minimization.  The conditional discord is written
\begin{align}
\Delta_{A;B|C} (\rho) &  =  d_{A;BC} (\rho) - d_{A;C} (\rho) \nonumber \\
& = S_{BC|\Pi^A} (\rho) -  S_{BC|A} (\rho)  \nonumber \\
&\qquad - S_{C|\Pi^A} (\rho) +  S_{C|A} (\rho) .
\end{align}
Using the fact that $ S_{BC|\Pi^A} (\rho)  = S_{BC|A} (\rho_{\Pi^A})  $ and $ S_{C|\Pi^A} (\rho)  = S_{C|A} (\rho_{\Pi^A}) $, we can evaluate each of the terms as
\begin{align}
S_{BC|\Pi^A} (\rho) & = S_{AB} ( \rho^{AB}_{\Pi^A} ) + S_{C} ( \rho^{C} ) - S_{A} ( \rho^{AB}_{\Pi^A} ) \nonumber \\
S_{BC|A} (\rho) & =  S_{AB} ( \rho^{AB} ) + S_{C} ( \rho^{C} )- S_{A} ( \rho^{AB} )  \nonumber \\
S_{C|\Pi^A} (\rho) & =  S_{C} ( \rho^{C} )\nonumber \\
 S_{C|A} (\rho) & = S_{C} ( \rho^{C} ) .
\end{align}
Substituting, we obtain
\begin{align}
\Delta_{A;B|C} (\rho) & = S_{B|A} ( \rho^{AB}_{\Pi^A} ) -  S_{B|A} ( \rho^{AB}) \nonumber \\
& = d_{A;B} (\rho) ,
\label{reductiondelta2}
\end{align}
which is the expression for the bipartite discord, without the minimization.  We note that one does not necessarily have the same optimal measurements in the tripartite discord in Eq. (8) of the main text.

\section{Properties of the monogamy $ \Delta_{A:B:C} $}

In this section, we describe some properties of the conditional discord  
$ \Delta_{A:B:C} $, as defined in Eq. (13) of the main text.


Here we show that for bipartite states $ \rho = \rho^{AB} \otimes \rho^C $, we have $ \Delta_{A:B:C}  = 0 $.  We know from (\ref{reductiondelta2}) that 
\begin{align}
\Delta_{A;B|C} (\rho^{AB} \otimes \rho^C) = d_{A;B} (\rho^{AB}) .
\end{align}
Furthermore, 
\begin{align}
\Delta_{A;C|B} (\rho^{AB} \otimes \rho^C) & = d_{A;BC} (\rho^{AB} \otimes \rho^C) -
d_{A;B} (\rho^{AB} \otimes \rho^C)   \nonumber \\
& = 0 
\end{align}
and
\begin{align}
d_{A;BC}(\rho^{AB} \otimes \rho^C) & = d_{A;B}(\rho^{AB})
\end{align}
Then from Eq. (13) in the main text, it follows that $ \Delta_{A:B:C}  = 0 $.

\end{document}